# Short-Term Voltage Security Constrained UC to Prevent Trip Faults in High Wind Power-Penetrated Power Systems

Yinhong Lin, *Student Member, IEEE*, Bin Wang, *Member IEEE*, Qinglai Guo, *Senior Member IEEE*, Haotian Zhao, *Member IEEE* and Hongbin Sun, *Fellow, IEEE*

*Abstract*—For high wind power-penetrated power systems, the multiple renewable energy station short-circuit ratio (MRSCR) is often insufficient due to weak grid structures. Additionally, transient voltage sag/overvoltage issues may cause trip faults of wind turbines (WTs). Due to the time delay in WTs' controllers, it is difficult for WTs alone to meet the reactive power demands in different stages of the transient process. Some synchronous machines (SMs) must be retained through unit commitment (UC) scheduling to improve MRSCR and prevent trip faults of WTs. The MRSCR and short-term voltage security constrained-UC model is a mixed integer nonlinear programming (MINLP) problem with differential algebraic equations (DAEs) and symbolic matrix inversion, which is intractable to solve. Based on the dynamic characteristics of different devices, the original model is simplified as a general MINLP model without DAEs. Then, generalized Benders decomposition is applied to improve the solution efficiency. The relaxed MRSCR constraints are formulated in the master problem to improve the convergence, and the precise MRSCR constraints are formulated in the subproblems to consider the impact of voltage profiles. Case studies based on several benchmark systems and a provincial power grid verify the validity and efficiency of the proposed method.

## NOMENCLATURE

*A. Acronyms*

| | |
|---|---|
| AVR | Automatic voltage regulator |
| DAE | Differential algebraic equation |
| GBD | Generalized Benders decomposition |
| HVDC | High-voltage direct current |
| HWPPS | High wind power-penetrated power system |
| IVS | Instantaneous voltage support |
| MIL | Mixed integer linear |
| MILP | Mixed integer linear programming |
| MINLP | Mixed integer nonlinear programming |
| MRSCR | Multiple renewable energy station short-circuit ratio |
| NLP | Nonlinear programming |
| PCC | Point of common coupling |
| SM | Synchronous machine |
| SCC | Short-circuit capacity |
| SOC | Second-order cone |
| SVSC | Short-term voltage security constraint |
| TOV | Transient overvoltage |
| TVS | Transient voltage sag |
| UC | Unit commitment |
| WF | Wind farm |
| WT | Wind turbine |

*B. Superscripts and subscripts of variables*

| | |
|---|---|
| $(\cdot)^{t,s}$ | Variable in period t under fault s |
| $(\cdot)^{t,0}$ | Variable in period t in steady states |
| $(\cdot)^{t}$ | Variable in period t |
| $(\cdot)_x / (\cdot)_y / (\cdot)_d / (\cdot)_q$ | x/y/d/q axis component of a variable |
| $(\cdot)^{k}$ | Value in the kth iteration |
| $(\cdot)'' / (\cdot)'$ | Variable in the subtransient/transient process |
| $(\cdot)_{min} / (\cdot)_{max}$ | Minimum/maximum value of a variable |
| $(\cdot)_g / (\cdot)_w / (\cdot)_l$ | Variable related to synchronous generator g/wind farm w/capacitor c/load l |
| $d(\cdot)$ | Diagonalization |

*C. Sets and parameters*

| | |
|---|---|
| $G / B$ | Real/imaginary part of node admittance matrix including load equivalent admittance |
| $S_N$ | Equipment capacity |
| $S_g / S_w / S_{flt} / S_T$ | Set of generator/wind farm/fault/time period |
| $V_N / I_N$ | Rated voltage/current |
| $x_g / x_w / x_l$ | Reactance of synchronous machines/wind turbines/transmission lines |
| $Y_0$ | Node admittance matrix |

*D. Variables*

| | |
|---|---|
| $C_P / C_U / C_D$ | Power generation cost/startup cost/shutdown cost of a synchronous machine |
| $I_{eq}$ | Potential equivalent current in synchronous machine |

Corresponding authors: Hongbin Sun (shb@tsinghua.edu.cn), Qinglai Guo (guoqinglai@tsinghua.edu.cn).
This work was supported in part by the Science and Technology Project of State Grid under Grant 5100-202199515A-0-5-ZN.

Y. Lin, B. Wang, Q. Guo, H. Zhao and H. Sun are with the State Key Laboratory of Power Systems, Department of Electrical Engineering, Tsinghua University, Beijing, 100084, China.

| | |
|---|---|
| $k_p / k_q$ | Active/reactive current loop gain of a wind turbine under low voltage ride through control |
| $P_{pre} / P_{cur}$ | Wind power forecast/reduction value |
| $PR_D^t / RU_g$ | Up/down active power reserve of a system |
| $RD_g / RU_g$ | Down/up ramp rate of a synchronous machine |
| $T_{on} / T_{off}$ | Minimum startup/shutdown time of a synchronous machine |
| $V / I / P / Q$ | Voltage/current/active power/reactive power |
| $V_{LVRT,th} / V_{HVRT,th}$ | Wind farm low voltage/high voltage ride through limit |
| $V_r$ | Voltage component generated by wind power |
| $y_g$ | On/off status of a synchronous machine |
| $Z / Y$ | Node impedance/admittance matrix including unit reactance |

I. INTRODUCTION

To achieve carbon neutrality, the capacity of renewable energy is rapidly improving in many countries. During 2022-2027, the installed capacity of renewable energy will reach 425 GW in Europe and 280 GW in the United States, and China will achieve 1200 GW of total installed capacity of wind and solar photovoltaic energy [1]. High penetration of renewable energy has become one of the features of modern power systems, which poses new challenges to the security of power grids, such as voltage security [2]. In high wind power-penetrated power systems (HWPPSs), a large number of synchronous machines (SMs) are replaced by wind turbines (WTs). Wind farm (WF) stations are generally located in grids with weak grid structures. Hence, the short-circuit capacity (SCC) of such a system is insufficient, limiting the maximum wind power penetration. To measure the system strength, the multiple renewable energy station short-circuit ratio (MRSCR) has been proposed for practical applications [3-4]. It is recommended that the MRSCR at the point of common coupling (PCC) be greater than 3.0 [5]. In addition, since a large number of inverter-based generators are installed in HWPPSs, the occurrence of short-circuit faults in AC systems and commutation failures of high-voltage direct current (HVDC) systems may lead to transient voltage sag/overvoltage (TVS/TOV) [6-7], which may cause cascading trip faults of WTs and result in wind power loss [8-9].

To eliminate the violation of MRSCR and short-term voltage security constraints (SVSCs), research has focused on planning [10-11], evaluation [12-13], unit commitment (UC) [14-16], network reconfiguration [14], and WT control strategies [17]. Due to the high cost of re-planning for wind farm stations during operation, a promising approach is to pursue maximization of the wind power consumption based on UC models considering MRSCR and SVSCs. When a large number of WTs are connected to replace SMs, the system strength will be insufficient due to the weak ability of WTs to provide short-circuit currents [13]. In addition, a WT generally shows current source characteristics [15]. There is a time delay in the reactive power support that WTs can provide in response to a sudden voltage change [18], thus the reactive power demands in different stages cannot be simultaneously met with WTs alone when the system voltage first decreases and then increases rapidly under faults [19]. Thus, it is crucial to retain some SMs to improve MRSCR and prevent trip faults of WTs. In addition, the WT control parameters have a great impact on the TVS/TOV issues [20], indicating that these parameters and the UC should be jointly optimized.

In existing research on UC problems, the UC has been optimized by adding short-circuit current constraints [14-15] or static voltage stability constraints [16] to the optimization model. Linear approximation or data-driven methods are generally used to establish constraints related to the SCC. However, when the operating conditions change greatly, the error of a linear model may not be acceptable. In addition, the inverter bus voltage amplitude difference has not been considered in these methods, which may be inaccurate for reactive power optimization [21]. In [22], the UC problem considering SVSCs of power grids is studied, and the voltage recovery performance is improved by optimizing the UC. However, for HWPPSs, they are prone to rapid changes in voltage which initially decreases and later increases under a fault [18]. This phenomenon is related to the instantaneous responses of the devices [23] and the responses of various controllers with different time scales [24-25]. In existing research, the voltage issues arising in different stages have rarely been considered simultaneously to jointly optimize the controllers and the UC.

Regarding the optimization of device controllers, existing studies have focused on improving the LVRT/HVRT control strategies for WTs [18][26] or HVDC strategies [7][27] to reduce the generation of excess reactive power and avoid TOV. Studies have focused on utilizing the reactive power regulation capability of each device through controller designs to suppress TOV, but have rarely considered the voltage issues arising in different stages [8-9]. Additionally, existing research has not addressed the coordination of different WTs at the system level to make full use of the reactive power support ability of all WTs during the transient process.

In summary, existing research has focused on the UC problem considering MRSCR constraints or WT/HVDC control strategies for suppressing TOV. However, different devices' support ability has not been fully utilized to prevent trip faults. For the coordinated optimization of UC and WT control parameters considering both MRSCR and SVSCs in HWPPSs, the challenges and countermeasures are listed as follows.

1) It is time-consuming to solve a reactive power optimization model with differential algebraic equations (DAEs). Some researchers have solved the model by evolution algorithms [28], but the performance depends on parameter settings [29]. Some have solved these models by discretization [30-31], but this approach faces the curse of dimensionality. Another common approach is trajectory sensitivity analysis [32-34]. However, the trajectory sensitivity calculation is time-consuming. To address this challenge, the dynamic characteristics of the devices can be utilized. Under large disturbances, due to flux linkage conservation [35] of the windings and the automatic voltage regulator (AVR) control time delay, an SM exhibits a constant voltage source. Due to the current source characteristics [15] and control time delay [18], a WT exhibits a constant current source. These dynamic characteristics can be used as boundary conditions to solve the network equations and obtain voltage amplitudes at critical moments.

2) The MRSCR constraints are difficult to formulate. These constraints involve nonlinear product terms and the inversion of a symbolic matrix [15]. Previous studies have generally used linear approximation [14] or the least squares method [15-16] to approximate constraints related to SCC. However, when the operating conditions change greatly, a linear model may be inaccurate. In addition, the impact of voltage profiles has not been considered in these methods [13]. In this paper, the MRSCR constraints are converted into two different forms. At first, the MRSCR constraints are relaxed by inequality scaling and converted into mixed integer linear constraints by discretization [36] and the Big M method [37]. The converted constraints are used to determine the UC and active power. Then, the MRSCR constraints considering precise voltage profiles are converted into second-order cones (SOCs), and are used for determining the reactive power and control parameters.

3) A large-scale mixed integer nonlinear programming (MINLP) model is difficult to solve directly. In day-ahead scheduling, multiple periods and multiple anticipated faults need to be considered in the optimization model. For such a large-scale MINLP model, it is difficult to obtain a feasible solution in an acceptable time. Thus, in [22][38], generalized Benders decomposition (GBD) is used to improve the solution efficiency. Inspired by [22], the original model can be decomposed into a mixed integer linear programming (MILP) master problem and several nonlinear programming (NLP) subproblems. The master problem optimizes UC and active power. The subproblems optimize reactive power and control parameters. Since the subproblems are decoupled, they can be solved by parallel computation.

The contributions of this paper are listed as follows.

1) The conflict on reactive power demands of TVS/TOV issues and the necessity of retaining some SMs are investigated based on theoretical derivations and time-domain simulations. Then, a method of simplifying DAEs into algebraic equation constraints is proposed. Based on the instantaneous characteristics and control time delay of SMs and WTs, the DAEs are simplified into algebraic equation constraints corresponding to the moments of fault occurrence, fault steady state, and fault clearance, thus avoiding the solution of DAEs.

2) A UC model considering MRSCR and SVSCs to prevent trip faults of WTs is formulated. The model is decomposed into an MILP master problem and NLP subproblems by GBD and can be solved effectively by commercial solvers. In addition, MRSCR constraints are formulated in the master problem and subproblems respectively to consider different factors. In the master problem, the constraints are relaxed into a mixed integer linear form to optimize UC and active power output. In the subproblems, the constraints considering voltage profiles are formulated into SOCs to optimize reactive power and control parameters.

The rest of this paper is organized as follows. In Section II, the mechanism of TVS/TOV is analyzed and then the voltage amplitudes at different moments are modeled by algebraic equations. In Section III, a UC model for multistage reactive power coordination considering MRSCR and SVSC is formulated. In Section VI, an efficient algorithm based on GBD is proposed. In Section V, based on several benchmark systems and a provincial power grid, verifications are carried out. Section VI draws the conclusions.

## II. MECHANISM ANALYSIS AND MODELING OF TRANSIENT VOLTAGE SAG/OVERVOLTAGE

### A. Mechanism analysis of reactive power responses

When short circuit faults occur in an AC system, the response of different devices and their controllers vary greatly [24-25]. For SMs, based on the flux linkage conservation theorem, some current will be induced in the excitation and damper windings at the moment of large disturbances. Hence, an SM exhibits the characteristic of a constant subtransient voltage source, and the instantaneous var output increases as the terminal voltage decreases, as shown in (1). Then, the excitation voltage is adjusted by an AVR. However, due to the time delay of exciters and other components, the impact of AVRs is minor. Since the current in the damper windings decreases much more rapidly than that in the excitation windings, in the fault steady state, an SM exhibits the characteristic of a constant transient voltage source. In contrast, due to the fast responses of current control loops, a WT exhibits the characteristics of a constant current source at the moment of large disturbances, and the instantaneous var output is proportional to the terminal voltage, as shown in (2). In the fault steady state, due to the LVRT control, the reactive current is injected into the system in accordance with the TVS, as shown in (3).

$$Q_g(t) = E_g''(t)V_g(t)/x_d'' \cos\delta(t) - V_g^2(t)/x_d'' \quad (1)$$

$$Q_w(t) = -1.5 V_{w,d}(t) I_{w,q}(t) \quad (2)$$

$$I_{w,q}^{LVRT}(t) = k_q I_N (V_{LVRT} - V(t)) \quad (3)$$

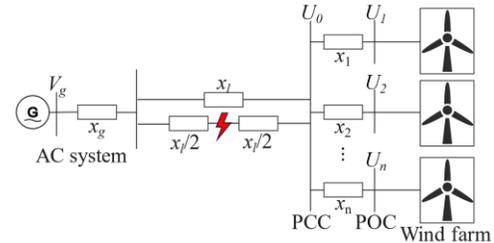

Fig. 1. High wind power-penetrated power system.

To verify the above analysis, an HWPPS is built in Fig. 1. The voltage and reactive power responses of different devices under large disturbances are shown in Fig. 2. Fig. 2(a) shows that when the short-circuit fault occurs or is cleared, the system voltage drops or rises instantaneously. Fig. 2(b) shows that the subtransient voltage of the SM remains unchanged at these moments. However, due to the voltage change at the terminals, there will be an instantaneous var response. Fig. 2(c) shows that the reactive current of the WT remains unchanged under large disturbances, whereas the reactive power changes proportionally to the voltage. From the moment of the fault occurrence to the moment of the fault clearance, there is an increase in the system voltage. During this process, the WT switches to LVRT control and injects reactive current in accordance with the TVS. For SMs, the transient voltage remains nearly unchanged, but the reactive power decreases due to the voltage rise. These phenomena verify the above conclusions.

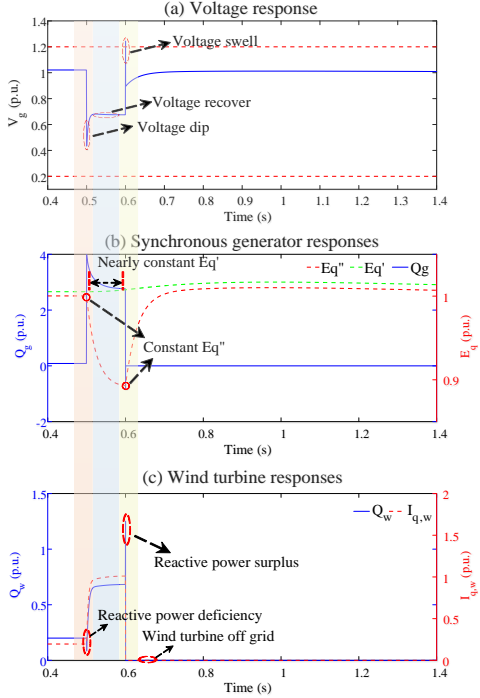

Fig. 2. Dynamic responses under a large disturbance.

### B. Modeling of transient voltage sag/overvoltage

To accurately characterize reactive power and voltage during the transient process, it is necessary to solve DAEs. However, the computation cost is high and it is difficult to apply for the optimal operation of large power systems. From the above analysis, it can be found that only certain moments need to be focused on, i.e., the moment of fault occurrence, the fault steady state, and the moment of fault clearance. Based on the mechanism analysis of reactive power responses in each stage, the boundary conditions for each stage are listed in Tab I.

TABLE I  BOUNDARY CONDITION FOR DIFFERENT STAGES DURING THE TRANSIENT PROCESS

| Stage | Synchronous machine | Wind turbine |
|---|---|---|
| Fault occurs($\tau_{1+}$) | $\begin{cases} I_{g,x}^{"t,s}(\tau_{1+}) = I_{g,x}^{"t,0} \\ I_{g,y}^{"t,s}(\tau_{1+}) = I_{g,y}^{"t,0} \end{cases}$ | $\begin{cases} I_{w,d}^{t,s}(\tau_{1+}) = I_{w,d}^{t,0} \\ I_{w,q}^{t,s}(\tau_{1+}) = I_{w,q}^{t,0} \end{cases}$ |
| Fault steady state($\tau_{2-}$) | $\begin{cases} I_{g,x}^{'t,s}(\tau_{2-}) = I_{g,x}^{'t,0} \\ I_{g,y}^{'t,s}(\tau_{2-}) = I_{g,y}^{'t,0} \end{cases}$ | $\begin{cases} I_{w,d}^{t,s}(\tau_{2-}) = k_{w,p}^t I_{w,d}^{t,0} \\ I_{w,q}^{t,s}(\tau_{2-}) = k_{w,q}^t I_N \\ (V_{LVRT,th} - V_w^{t,s}(\tau_{2-})) \end{cases}$ |
| Fault clears($\tau_{2+}$) | $\begin{cases} I_{g,x}^{"t,s}(\tau_{2+}) = I_{g,x}^{"t,s}(\tau_{2-}) \\ I_{g,y}^{"t,s}(\tau_{2+}) = I_{g,y}^{"t,s}(\tau_{2-}) \end{cases}$ | $\begin{cases} I_{w,d}^{t,s}(\tau_{2+}) = I_{w,d}^{t,s}(\tau_{2-}) \\ I_{w,q}^{t,s}(\tau_{2+}) = I_{w,q}^{t,s}(\tau_{2-}) \end{cases}$ |

where $I''$ and $I'$ represent the Norton equivalent current corresponding to subtransient and transient voltages respectively. Based on Tab I, the voltage amplitudes can be calculated directly by solving network equations [39].

To analyze the TVS/TOV issues in HWPPSs, based on the system in Fig. 1, the voltage amplitudes at PCC at the moments of the fault's occurrence $V_{flt}$ and clearance $V_{clr}$ are derived. For simplicity, it is assumed that WTs inject only reactive current under the fault. The constant internal voltage model is adopted for the SM. When more SMs are turned on in the system, it is represented by reducing the equivalent reactance $x_g$. When the wind power capacity is increased, it is represented by increasing injection current $I_q$. Note that these simplifications are adopted only for analyzing the mechanisms of TVS/TOV. An accurate model will be used in the following optimization model. The voltage amplitudes at the PCC before and after the fault clearance can be derived in (4)(5). When the number of wind farms $N_w$ is 1, the diagram is shown in Fig. 3.

$$V_{flt} = V_{LVRT,th} - \frac{V_{LVRT,th} - V_g x_l / (8x_g + 3x_l)}{1 + I_N k_q (N_w x_l (3x_g + x_l) / (8x_g + 3x_l) + x_w)} \quad (4)$$

$$V_{clr} = V_g + \frac{I_N k_q (N_w (x_g + x_l/2) + x_w)(V_{LVRT,th} - V_g x_l / (8x_g + 3x_l))}{1 + I_N k_q (N_w x_l (3x_g + x_l) / (8x_g + 3x_l) + x_w)} \quad (5)$$

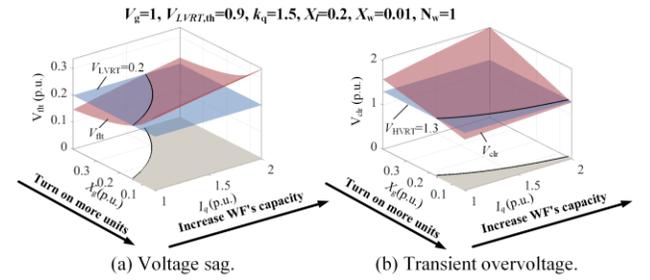

Fig. 3 The impacts of unit commitment and reactive current capacity of wind turbines on transient voltage sag/overvoltage.

Based on Fig. 3(a)(b), when fewer SMs are operating in the system, the instantaneous voltage support (IVS) ability of the system is weaker, and the TVS/TOV issues are more severe. Thus, WTs' trip faults are more likely to occur. Fig. 3(a) shows that increasing the reactive current injected from the WTs can mitigate the TVS issue. However, due to the time delay of the WT's reactive power control, Fig. 3(b) shows that the reactive current will be excessive when the fault is cleared, making the TOV issue more severe. Thus, TVS and TOV issues present a conflict in terms of the reactive power demands, and WTs alone cannot meet the demands of these issues simultaneously. Thus, to meet both the TVS and TOV security constraints, it is necessary to retain a certain number of SMs to provide IVS.

In summary, the following conclusions can be drawn:

1) At the moments of fault occurrence and clearance, an SM exhibits constant subtransient voltage characteristics. Its instantaneous var increases or decreases as the voltages drop or swell. In contrast, a WT exhibits constant current source characteristics. Its instantaneous var decreases or increases as the voltages drop or swell.

2) Due to the time delay of WTs' controllers, the current reference cannot follow a change in voltage immediately. When the voltage "first decreases then increases" rapidly under a fault, it is difficult for WTs alone to meet the reactive power demands of different stages simultaneously. Therefore, it is necessary to retain a certain number of SMs to provide IVS.

To promote wind power consumption, it is necessary to turn on as few units as possible while ensuring sufficient SCC and IVS. Since the dynamic voltage support possesses spatially distributed characteristics, it is necessary to jointly optimize UC and WT control parameters considering the power grid topology. An optimization model for this purpose is introduced in the next section.

## III. Unit Commitment Model for Multistage Reactive Power Coordination

### A. Original model

The complete UC model considering MRSCR and SVSCs is expressed as follows:

$$\min \sum_t C_{g,P}^t + C_{g,U}^t + C_{g,D}^t + MP_{w,cur}^t \quad (6)$$

$$\text{s.t. (7)-(31)}$$

The objective is to minimize the operation costs of the units within one day, including generation, startup, and shutdown costs. Additionally, the wind power curtailment is added to the objective as a penalty term, where $M$ is the penalty coefficient. The expressions for the objective and constraints are as follows. The piecewise linearization is applied to the generation costs.

*1) Objective function (generation/startup/shutdown costs)*

$$C_{g,P}^t = y_g^t f_n(\underline{P_g}) + \sum_{m=1}^{M_n} F_{g,m} P_{g,m}^t$$

$$C_{g,U}^t \geq C_{g,su}(y_g^t - y_g^{t-1}), C_{g,U}^t \geq 0 \quad (7)$$

$$C_{g,D}^t \geq C_{g,sd}(y_g^{t-1} - y_g^t), C_{g,D}^t \geq 0, \forall g \in S_g, \forall t \in S_T$$

where $F_{g,m}$ is the slope of unit $g$ in segment $m$.

*2) Piecewise linear active power constraints*

$$P_g^{t,0} = y_g^t P_{g,\min} + \sum_{m=1}^{M_n} P_{g,m}^t, 0 \leq P_{g,m}^t \leq P_{g,m} - P_{g,m-1} \quad (8)$$

$$P_{g,0} = P_{g,\min}, P_{g,M_g} = P_{g,\max}$$

where $P_{g,m}$ is the maximum output of unit $g$ in segment $m$.

*3) Minimum startup/shutdown times*

$$y_g^h \geq y_g^t - y_g^{t-1} \quad (h = t, t+1, ... t + T_{g,on} - 1) \quad (9)$$

$$1 - y_g^h \geq y_g^{t-1} - y_g^t \quad (h = t, t+1, ... t + T_{g,off} - 1), \forall g \in S_g, \forall t \in S_T \quad (10)$$

*4) Ramp rate constraints*

$$-RD_g \leq P_g^{t,0} - P_g^{t-1,0} \leq RU_g, \forall g \in S_g, \forall t \in S_T \quad (11)$$

*5) System reserve capacity constraints*

$$\sum_{g \in S_g} (y_g^t P_{g,\max} - P_g^{t,0}) \geq PR_U^t, \forall t \in S_T \quad (12)$$

$$\sum_{g \in S_g} (P_g^{t,0} - y_g^t P_{g,\min}) \geq PR_D^t, \forall t \in S_T \quad (13)$$

*6) Current and power equations for synchronous machines*

$$\begin{bmatrix} I_{g,x}^{t,s}(\tau) \\ I_{g,y}^{t,s}(\tau) \end{bmatrix} = -\begin{bmatrix} 0 & y_g^t / x_{eq} \\ -y_g^t / x_{eq} & 0 \end{bmatrix}\begin{bmatrix} V_{g,x}^{t,s}(\tau) \\ V_{g,y}^{t,s}(\tau) \end{bmatrix} + \begin{bmatrix} I_{g,x}^{eq,t,s}(\tau) \\ I_{g,y}^{eq,t,s}(\tau) \end{bmatrix} \quad (14)$$

$$P_g^{t,s}(\tau) = V_{g,x}^{t,s}(\tau) I_{g,x}^{t,s}(\tau) + V_{g,y}^{t,s}(\tau) I_{g,y}^{t,s}(\tau) \quad (15)$$

$$Q_g^{t,s}(\tau) = V_{g,y}^{t,s}(\tau) I_{g,x}^{t,s}(\tau) - V_{g,x}^{t,s}(\tau) I_{g,y}^{t,s}(\tau) \quad (16)$$

$$\forall g \in S_g, \forall s \in S_{flt}, \forall t \in S_T$$

where '$eq$' indicates variables during the subtransient or transient process.

*7) Power equations for wind farms*

$$P_w^{t,s}(\tau) = 1.5 V_w^{t,s}(\tau) I_{w,d}^{t,s}(\tau) = V_{w,x}^{t,s}(\tau) I_{w,x}^{t,s}(\tau) + V_{w,y}^{t,s}(\tau) I_{w,y}^{t,s}(\tau) \quad (17)$$

$$Q_w^{t,s}(\tau) = -1.5 V_w^{t,s}(\tau) I_{w,q}^{t,s}(\tau) = V_{w,y}^{t,s}(\tau) I_{w,x}^{t,s}(\tau) - V_{w,x}^{t,s}(\tau) I_{w,y}^{t,s}(\tau) \quad (18)$$

$$\forall w \in S_w, \forall t \in S_T, \forall s \in S_{flt}$$

*8) Network equations (prefault, during fault, and postfault)*

$$\begin{bmatrix} I_x^{t,s}(\tau) \\ I_y^{t,s}(\tau) \end{bmatrix} = \begin{bmatrix} G^s & -B^s \\ B^s & G^s \end{bmatrix}\begin{bmatrix} V_x^{t,s}(\tau) \\ V_y^{t,s}(\tau) \end{bmatrix} \quad (19)$$

*9) Station voltage constraints*

$$V_{w,\min} \leq V_w^{t,0} \leq V_{w,\max}, \forall w \in S_w, \forall t \in S_T \quad (20)$$

*10) Short circuit ratio constraints*

$$\frac{\left| V_{N,w} V_w^{t*} / Z_{w,w}^t \right|}{\sum_{j \in S_w} \left| (Z_{w,j}^t V_w^{t*}) / (Z_{w,w}^t V_j^{t*}) \right| P_j^{t,0}} \geq MRSCR_{th} \quad (21)$$

$$\mathbf{Z}^t = \left[\mathbf{Y}^t\right]^{-1} = \left[\mathbf{Y}_0 + diag\left(\mathbf{y}^t \cdot \mathbf{x}_d''\right)\right]^{-1}, \forall t \in S_T, \forall w \in S_w \quad (22)$$

*11) Voltage constraints at the fault occurrence*

$$V_w^{t,s}(\tau_{1+}) \geq V_{LVRT,th}, \forall w \in S_w, \forall t \in S_T, \forall s \in S_{flt} \quad (23)$$

*12) Voltage constraints at the fault steady state*

$$V_w^{t,s}(\tau_{2-}) \geq V_{LVRT,th}, \forall w \in S_w, \forall t \in S_T, \forall s \in S_{flt} \quad (24)$$

*13) Voltage constraints at the fault clearance*

$$V_w^{t,s}(\tau_{2+}) \leq V_{HVRT,th}, \forall w \in S_w, \forall t \in S_T, \forall s \in S_{flt} \quad (25)$$

*14) Generator capacity limits*

$$y_g^t P_{g,\min} \leq P_g^{t,0} \leq y_g^t P_{g,\max} \quad (26)$$

$$y_g^t Q_{g,\min} \leq Q_g^{t,0} \leq y_g^t Q_{g,\max}, \forall g \in S_g, \forall t \in S_T \quad (27)$$

*15) Wind turbine capacity limits*

$$0 \leq P_w^{t,0} \leq P_{w,pre}^t - P_{w,cur}^t \quad (28)$$

$$P_w^{t,s2} + Q_w^{t,s2} \leq S_{w,\max}^2, \forall w \in S_w, \forall t \in S_T, \forall s \in S_{flt} \quad (29)$$

*16) Current loop gains in low-voltage ride through*

$$0 \leq k_{w,p}^t \leq 0.5, \quad 0 \leq k_{w,q}^t \leq 3, \forall w \in S_w, \forall t \in S_T \quad (30)$$

*17) Differential equations of dynamic components*

$$d\mathbf{x}(\tau)/d\tau = f(\mathbf{u}, \mathbf{x}(\tau), \mathbf{y}(\tau)), \quad \tau \in [\tau_1, \tau_2] \quad (31)$$

where $\mathbf{x}$ represents the state variables, such as SMs' subtransient voltages, WTs' injection currents, etc. $\mathbf{y}$ represents algebraic variables, such as voltages, reactive powers, etc. $\mathbf{u}$ represents control variables, such as UC, WT parameters, etc.

The model above includes DAEs (31) that are intractable to solve. Therefore, the model is further processed as follows.

### B. Simplification of DAEs

For convenience, the compact form of the model is as follows.

$$\min C(\mathbf{u}) \quad (32)$$

$$\text{s.t. } \mathbf{G}(\mathbf{u}, \mathbf{x}_0, \mathbf{y}_0) = \mathbf{0} \quad (33)$$

$$\underline{\mathbf{H}_1} \leq \mathbf{H}_1(\mathbf{u}, \mathbf{x}_0, \mathbf{y}_0) \leq \overline{\mathbf{H}_1} \quad (34)$$

$$d\mathbf{x}(\tau)/d\tau = f(\mathbf{u}, \mathbf{x}(\tau), \mathbf{y}(\tau)) \quad (35)$$

$$\mathbf{g}(\mathbf{x}(\tau), \mathbf{y}(\tau), \mathbf{u}) = \mathbf{0}, \tau \in [\tau_{1-}, \tau_{2+}] \quad (36)$$

$$\underline{\mathbf{H}_2} \leq \mathbf{H}_2(\mathbf{x}(\tau), \mathbf{y}(\tau)) \leq \overline{\mathbf{H}_2} \quad (37)$$

(33) and (34) represent power flow equations, security constraints in steady states, and feasible regions of control variables. (35) and (36) are DAEs that characterize the dynamic responses of the system. (37) represents SVSCs. For (35) and (36), the common method is to remove them from the model and solve them by time-domain simulations. Then, some linear cuts are introduced back to the model based on trajectory sensitivity [32-34]. However, this method may suffer from low efficiency. Nevertheless, since STVS issues in HWPPSs generally occur either at the moments of fault occurrence ($\tau_{1+}$), in the fault steady state ($\tau_{2-}$), or at the moment of fault clearance ($\tau_{2+}$), it is not necessary to calculate the complete responses. Instead, if the values or the algebraic equations for state variable $x(\tau)$ at certain moments can be obtained in advance, then $y(\tau)$ at the corresponding moments can be directly obtained by solving the network equation (36). Tab. I shows the algebraic equations for various state variables $x(\tau)$ in different stages, which can be rewritten in a compact form as shown in (38). By replacing (35) with (38), the original model containing DAEs is converted into a general MINLP model.

$$F(x(\tau), y(\tau), x_0) = 0, \tau \in \{\tau_{1+}, \tau_{2-}, \tau_{2+}\} \quad (38)$$

## IV. EFFICIENT SOLUTION METHOD BASED ON GENERALIZED BENDERS DECOMPOSITION

After the simplification in Section III-B, the original model is converted into a general MINLP model. However, when such a model is applied to large-scale power grids, since the mathematical form is MINLP, it is still difficult to solve the model directly in an acceptable time. Thus, an efficient solution approach is proposed to support the application of the SVSC-UC in real power systems.

Inspired by [22], GBD algorithm is applied to decompose the original model into an MILP master problem and several NLP subproblems. However, the MRSCR constraints (21) involve the inversion of the symbolic matrix $Y$ containing binary variables (22), which is difficult to express explicitly. Moreover, the denominator in (21) contains a nonlinear product term of three continuous variables. To overcome the error caused by linear approximation method and the neglect of voltage profiles [13-14][35], the MRSCR constraints are converted into different forms. First, the MRSCR constraints are relaxed into an MIL form in the master problem to improve the convergence and determine the UC. Then, based on fixed UC, the precise MRSCR constraints considering voltage profiles are formulated into SOCs in the subproblems. The overall framework of the proposed method is shown in Fig. 4. Once the integer variables have been determined in the master problem, the Hessian matrices of the subproblems are constant, so the interior point method [40] can be used to solve the subproblems efficiently. Parallel computation is also applied to the solution process.

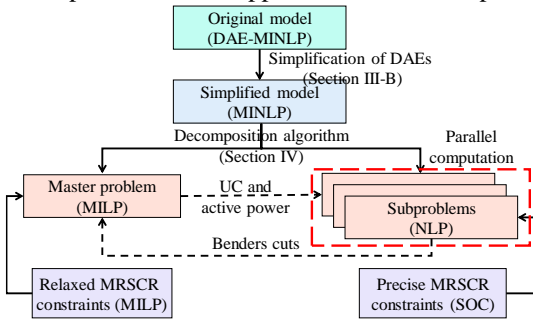

Fig. 4. Framework of the proposed method.

### 1) Master problem

Considering active power balance and the relaxed MRSCR constraints, the master problem minimizes the total costs by optimizing the UC and the active power. The complete model is as follows.

$$\min \sum_t C_{g,P}^t + C_{g,U}^t + C_{g,D}^t + MP_{w,cur}^t \quad (39)$$

s.t. (7)-(13),(26),(28)

In addition, the following constraints need to be added.
*a) Active power balance constraints*

$$\sum_{g \in S_g} P_g^{t,0} + \sum_{w \in S_w} P_w^{t,0} = \sum_{l \in S_l} P_l^t \quad (40)$$

*b) Feasible cuts*

$$PO^{t,k} + \lambda_{y_g^t}^k (y_g^t - y_g^{t,k}) + \lambda_{P_g^{t,0}}^k (P_g^{t,0} - P_g^{t,0,k}) + \lambda_{P_w^{t,0}}^k (P_w^{t,0} - P_w^{t,0,k}) \leq 0 \quad (41)$$

where $\lambda_{y_g^t}^k$, $\lambda_{P_g^{t,0}}^k$, and $\lambda_{P_w^{t,0}}^k$ are Lagrange multipliers corresponding to constraint (46).

*c) Relaxed form of MRSCR constraints*

The numerator and denominator in (21) are multiplied by the self impedance $Z_{w,w}$, and the inequality is scaled according to the voltage limit [$V_{min}$, $V_{max}$]. Then (42) is yielded. Since product terms on the right side of (42) is bilinear, binary discretization is applied [36] to the wind power curtailment in (42). Using the Big M method [37], (42) is converted into a linear form in (43).

$$V_{N,w} / MRSCR_{th} \geq Z_{w,w}^t P_w^t / V_{w,max} + \sum_{j \in S_w} Z_{w,j}^t P_j^t V_{w,min} / V_{w,max}^2 \quad (42)$$

$$w \in S_w, t \in S_T$$

$$Z_{w,j}^t (P_{j,pre}^t - P_{j,cur}^t) = Z_{w,j}^t P_{j,pre}^t - \sum_{n \in S_{int}} 2^{n-1} p_{w,j}^{t,n} \Delta P_{j,cur}^t$$

$$-M(1 - v_j^{t,n}) + Z_{w,j}^t \leq p_{w,j}^{t,n} \leq M(1 - v_j^{t,n}) + Z_{w,j}^t \quad (43)$$

$$-M v_j^{t,n} \leq p_{w,j}^{t,n} \leq M v_j^{t,n}, w \in S_w, t \in S_T$$

Since $Z$ is the inverse of $Y$ in (22), by applying the Big M method, (44) can be obtained.

$$\sum_{k \in S_b} Z_{wk}^t Y_{kg,0}^t + v_{wg}^t / x_{g,d}^{''} = \begin{cases} 1 & w = g \\ 0 & w \neq g \end{cases}$$

$$\sum_{k \in S_b} Z_{wk}^t Y_{kg,0} = \begin{cases} 1 & w = g \\ 0 & w \neq g \end{cases} \quad (44)$$

$$-M(1 - y_g^t) + Z_{wg}^t \leq v_{wg}^t \leq M(1 - y_g^t) + Z_{wg}^t$$

$$-M y_g^t \leq v_{wg}^t \leq M y_g^t, w \in S_w, g \in S_g, g \in \mathbb{C}_U S_g$$

### 2) Subproblems

The subproblems optimize the reactive power output of the units and the parameters of the WTs considering the precise MRSCR constraints and SVSCs under fixed UC status. Meanwhile, to ensure that the subproblems are feasible, the security constraints are relaxed, and a penalty term is added to the objective function. The complete model is as follows.

$$\min PO^t = \sum_{w \in S_w} \xi_{w,SCR}^t + \sum_{s \in S_{flt}} \sum_{w \in S_w} (\xi_{w,flt}^{t,s} + \xi_{w,fss}^{t,s} + \xi_{w,clr}^{t,s}) \quad (45)$$

s.t. (14)-(20),(27),(29),(30),(38) $\forall t \in S_T$

In addition, the following constraints need to be added.
*a) Coupling constraints*

$$y_g^t = y_g^{t,k} : \lambda_{y_g^t}^k, \; P_g^{t,0} = P_g^{t,0,k} : \lambda_{P_g^{t,0}}^k, \; P_w^{t,0} = P_w^{t,0,k} : \lambda_{P_w^{t,0}}^k \quad (46)$$

$$\forall t \in S_T, \forall g \in S_g, \forall w \in S_w, \; \forall c \in S_c$$

*b) Precise form of MRSCR constraints*

After determining UC, to consider the voltage profile, the equivalent definition of MRSCR in (47) is adopted [41]. It is converted to an SOC in (48). The voltage component in (48) is generated by the WTs' injection power and meets (49). These constraints are linear once the UC status has been determined.

$$MRSCR_w^t = V_{N,w}/V_{w,r}^t \ge MRSCR_{th} \quad (47)$$

$$V_{w,x,r}^{t2} + V_{w,y,r}^{t2} \le (V_{N,w}/MRSCR_{th})^2 + \xi_{w,SCR}^t, \xi_{w,SCR}^t \ge 0 \quad (48)$$

$$\begin{bmatrix} I_{w,x}^{t,0} \\ I_{w,y}^{t,0} \end{bmatrix} = \begin{bmatrix} G^0 & -B^0 \\ B^0 & G^0 \end{bmatrix} \begin{bmatrix} V_{w,x,r}^t \\ V_{w,y,r}^t \end{bmatrix} + \begin{bmatrix} 0 & y_g^t/x_{g,d}^{''} \\ -y_g^t/x_{g,d}^{''} & 0 \end{bmatrix} \begin{bmatrix} V_{w,x,r}^t \\ V_{w,y,r}^t \end{bmatrix} \quad (49)$$

*c) Voltage constraints at fault occurrence*

$$V_w^{t,s}(\tau_{1+}) + \xi_{w,flt}^{t,s} \ge V_{LVRT,th}, \xi_{w,flt}^{t,s} \ge 0 \quad (50)$$

*d) Voltage constraints in the fault steady state*

$$V_w^{t,s}(\tau_{2-}) + \xi_{w,fss}^{t,s} \ge V_{LVRT,th}, \xi_{w,fss}^{t,s} \ge 0 \quad (51)$$

*e) Voltage constraints at fault clearance*

$$V_w^{t,s}(\tau_{2+}) - \xi_{w,clr}^{t,s} \le V_{HVRT,th}, \xi_{w,clr}^{t,s} \ge 0 \quad (52)$$

where $\xi_{SCR}^t$, $\xi_{w,flt}^{t,s}$, $\xi_{w,fss}^{t,s}$, and $\xi_{w,clr}^{t,s}$ are the slack variables of constraints (48)(50)(51)(52), respectively.

The solution process is as follows. The master problem and subproblems are solved alternately. When the objectives of the subproblems are less than a certain threshold, such as $10^{-4}$, the solution is considered to satisfy all constraints in the original model, and the iterative process terminates. Otherwise, the Benders cuts in (41) are sent to the master problem. Since different subproblems can be solved in parallel, this algorithm can handle constraints for multiple periods efficiently.

## V. CASE STUDY

In this section, the computation is conducted using a Legion laptop with an Intel i7-9750H CPU and 16 GB of RAM. The MILP master problem is solved with GUROBI 9.1.1 [42]. The NLP subproblems are solved with IPOPT 3.12.3 [43]. The first test system is a modified version of IEEE 39-bus system [44-45], in which WFs are installed at buses 11, 14, 19, and 40, as shown in Fig. 5. For the IEEE 118-bus system, the locations and capacities of WFs can be found in [46]. The actual power grid considered in this section includes 1852 buses, 50 WFs and 50 SMs. The voltage thresholds for WTs are set to 0.2 and 1.2 [47].

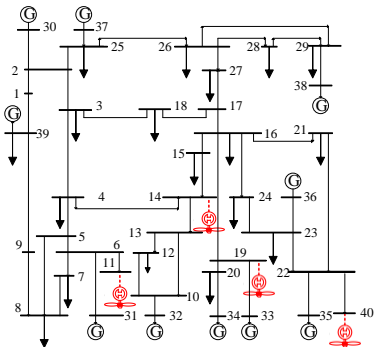

Fig. 5. Modified IEEE 39-bus system.

*A. Accuracy of the transient voltage modeling*

To test the calculation accuracy of transient voltage at critical moments, a comparison between the proposed model calculation results with the time-domain simulation results based on IEEE 39-bus system is shown in Fig. 6. At the moments of the fault occurrence, fault steady state and fault clearance, the maximum error of the voltages does not exceed 7.9408e-04(p.u.), 9.3564e-05(p.u.), 0.0079(p.u.), respectively. Thus, the simplified model can calculate the bus voltage at each critical moment efficiently with sufficient accuracy.

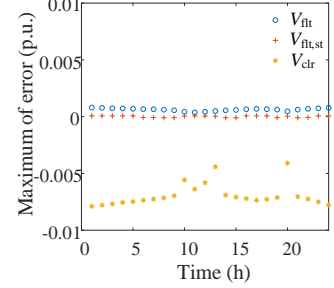

Fig. 6. Error analysis of voltage amplitudes at different moments.

*B. Validity of the proposed model*

To analyze the effects of considering MRSCR and TVS/TOV constraints, both UC and SVSC-UC models are solved, and the results are shown in Tab II. It shows that the operation costs based on SVSC-UC models are increased compared with the costs based on UC models.

TABLE II OPERATION COST IN DIFFERENT SCENARIOS.

| System | Overall cost ($) | Percentage uplift |
|---|---|---|
| IEEE 39 | 371,220 | 1.1% |
| IEEE 118 | 438,810 | 0.6% |
| Real system | 4,773,900 | 0.7% |

Take the IEEE39-bus system as an example. The operation cost of SVSC-UC results is increased by 1.1% compared with UC results, which is within the acceptable range. This is because more units are turned on after considering the MRSCR and SVSCs. The results are shown in Fig. 7. It can be seen that unit 36 near WF 40 is turned on, while the more remote units 34 or 37 are turned off, thereby enhancing the SCC and IVS in the area. Meanwhile, the reactive current gains of the WTs on bus 40 are reduced to avoid excess reactive current injection.

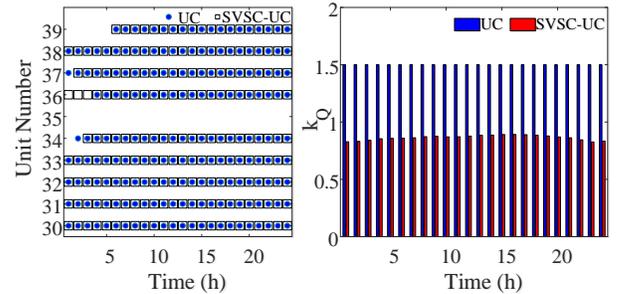

(a) Unit commitment   (b) Control parameters
Fig. 7. UC and control parameters of IEEE 39-bus system.

For WF 40, the MRSCR throughout one day based on UC and SVSC-UC models are compared in Fig. 8(a). It can be observed that the MRSCR is closely related to both wind power and the number of units turned on. When the wind power is large and only a few units are turned on, the MRSCR is more likely to be violated. Take the first period as an example. In UC results, the wind power integration is large during the first period, and only 6 SMs are turned on, among which units 30, 37, and 38 are far from WF 40. Consequently, the MRSCR cannot meet the requirement. In contrast, in SVSC-UC results, unit 36 near WF 40 is turned on, and remote unit 37 is turned off. As a result, the SCC of WF 40 is increased by nearly 500 MVA, and the violation of MRSCR is eliminated. By continuously increasing the wind power integration and reducing the active power of SMs, the MRSCR of WF 40 is obtained, as shown in Fig. 8(b). The maximum allowable wind power in the UC results is 931 MW, while the maximum allowable wind power in the SVSC-UC results is 1758 MW. The increase in wind power is nearly 88.7%. This shows that the proposed model can promote wind power consumption.

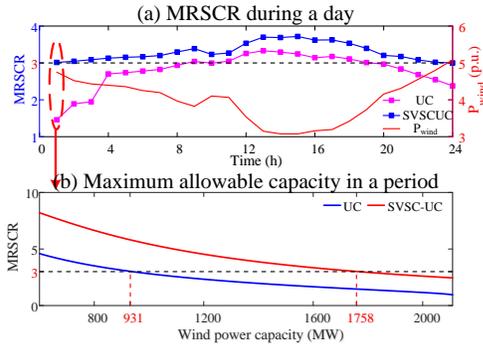

Fig. 8. MRSCR and maximum capacity based on UC and SVSC-UC models.

Based on time-domain simulations, the voltage security of different WFs in different periods is shown in Fig. 9. Fig. 9(a) shows the WFs' security in different periods. In the UC results, for the first 5 periods, the reactive power outputs of WTs and SMs in different stages of the transient process are not fully coordinated. The WTs on bus 14 have a high risk of tripping under the fault due to TVS/TOV. In addition, in the first period, all the WTs face a risk of tripping.

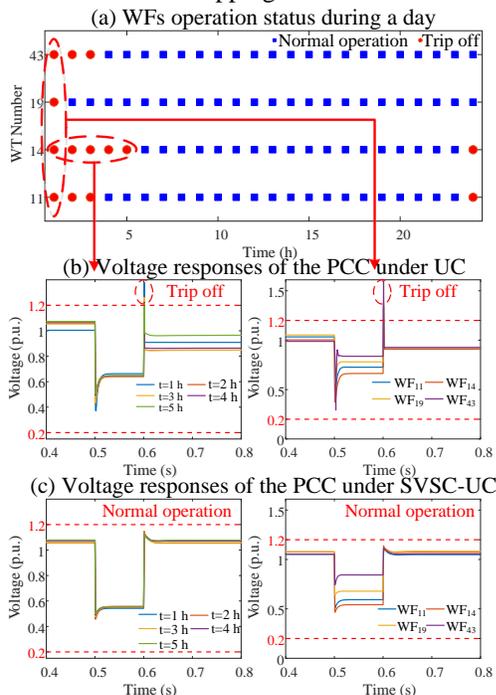

Fig. 9. Voltage responses at PCCs during different periods in a day.

The voltage responses are shown in Fig. 9(b). In UC results, the minimum and maximum voltage amplitudes of bus 14 across different periods are 0.37 and 1.38, respectively, which violate the TOV constraints. In SVSC-UC results, since unit 36 near WF 14 is turned on, and the reactive power capacities of SMs and WTs are fully coordinated, the TVS/TOV issues are significantly mitigated, and the voltage amplitudes at the moments of the fault occurrence and clearance are 0.45 and 1.15, respectively. Therefore, the WTs can operate normally.

To further analyze the reactive power coordination principle, the var output of different devices under the fault is investigated. Fig. 10(a) shows the total reactive power output of SMs and WTs at different moments in the transient process. In the SVSC-UC results, the initial reactive power output of the SMs and WFs are decreased and increased by 668 MVar and 1238 MVar, respectively. The instantaneous var absorbing ability of SMs is improved when the fault is cleared. At the moment of short circuit fault occurrence, the total instantaneous var output increased by 437 MVar due to the support of WTs, and the voltage sag extent decreases by nearly 0.09 (p.u.). During the steady state under the fault, since the control parameters of the WTs are decreased in advance, the total reactive power injected by different WFs under the LVRT control decreased by 1022 MVar, avoiding excess reactive power at the moment of the fault clearance. When the fault is cleared, the excess reactive power of the WTs due to the control delay has decreased by nearly 855 MVar, and the SMs absorb nearly 585.6 MVar reactive power. Thus, the TOV is reduced by 0.24 (p.u.), and the voltage is below 1.2, avoiding tripping of the WTs. This shows that the SVSC-UC model can effectively coordinate the dynamic voltage support capability at different moments during the transient process and maintain the security of the system.

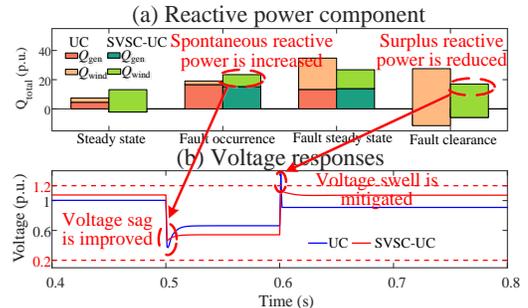

Fig. 10. Voltage responses and reactive power decomposition.

### C. Computation efficiency of the algorithm

To test the computation efficiency of the algorithm, SVSC-UC models are formulated based on Modified IEEE 39, IEEE 118, and a real power system. The computation time is shown in Tab III. Note that only a few contingencies may cause the issues of TVS/TOV. Thus, a severe fault set can be obtained through contingency screening before optimization [48], thereby reducing the computation burden of the subproblems. Besides, since the coupling constraints are considered in the master problem, the subproblems are decoupled from each other. Thus, parallel computation can be applied to accelerate the solution. With the above effort, the solution of the master problem becomes the bottleneck, and its computation time accounts for nearly 84%-98% of the total time.

Take IEEE 39-bus system as an example. The master problem is an MILP problem with 720 integers. The total computation time is 60 s. The subproblems are quadratically constrained quadratic programming. Thus, the Hessian matrices of the subproblems are constant. When solving the subproblems with the interior point method, the Hessian matrices need to be calculated only once. The average computation time for each subproblem is 11.2 s. Finally, the solution is obtained after 8 iterations, and the total computation time is 71.2 s. The solver Baron is also applied to solve the model [49]. However, it can not obtain a feasible solution after solving for more than 5 h, which shows that the proposed algorithm is efficient.

TABLE III COMPUTATION TIME OF DIFFERENT SCENARIOS.

| System | Proposed method ||||  Baron solver/s |
|---|---|---|---|---|---|
| | Master problem/s | Subproblem/s | Iterations | Total/s | |
| IEEE 39 | 60 | 11.2 | 8 | 71.2 | >18000 |
| IEEE 118 | 547 | 28.5 | 5 | 575.5 | >18000 |
| Real system (1852 buses) | 4822.8 | 107.6 | 4 | 4930.4 | >18000 |

As the scale of the power grids become larger, the computation time of the master and slave problem increases. For a provincial power grid with 1852 buses, when 20 anticipated faults are considered, the computation time of the master problem and subproblem is 4822.8 s and 107.6 s, respectively. After 4 iterations, the solution is obtained, and the total computation time is about 1.37 h, which is still within the acceptable range and can meet the requirements of the day-ahead scheduling.

*D. Analysis of influencing factors*

The results of the model are affected by several key factors. Take the TOV threshold and wind power penetration as examples. The analysis is based on the IEEE 39-bus system.

*1) Transient overvoltage threshold*

As shown in Fig. 11(a), as the threshold of TOV decreases, the number of units turned on increases, and the IVS at the moment of large disturbances is more sufficient. However, the operation cost also increases. When the threshold of TOV is increased past a certain extent, the operation cost of the system is nearly unchanged. This is because the MRSCR and other constraints become the main factors limiting wind power consumption, while the SVSCs are not activated. Hence, a certain number of SMs must be reserved to meet other constraints, so the cost remains nearly unchanged. In real applications, the threshold should be determined according to the overvoltage-withstanding ability of the WTs. Therefore, it is necessary to comprehensively consider the system operation cost and the WT cost to select the appropriate devices.

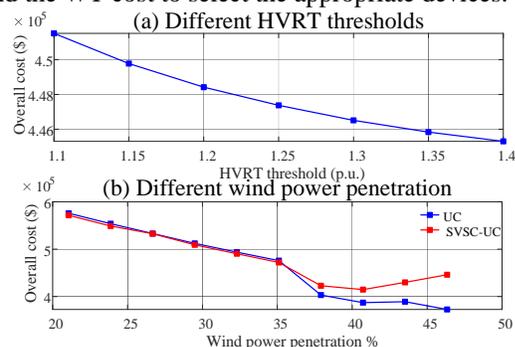

Fig. 11. Overall cost in different scenarios.

*2) Wind power penetration*

As shown in Fig. 11(b), as the wind power penetration increases and the SM power decreases, the operation cost decreases. However, when the penetration exceeds 37%, the MRSCR or TVS/TOV constraints tend to be violated. To meet the constraints, additional units need to be turned on to provide SCC and IVS, and the operation costs are increased. Since the active power of the SMs is subject to a lower bound, when the penetration of wind power reaches a certain level, most SMs in the system have reached the lower bound and no more units can be turned on. In this case, MRSCR violations need to be eliminated through wind power curtailment, which makes it difficult to further improve wind power penetration. To further improve the maximum allowable wind power capacity, the SCC and IVS of the system can be enhanced by improving the grid structure or installing synchronous condensers in the planning stage.

## VI. CONCLUSION

Through mechanism analysis and time-domain simulation of the reactive power responses of SMs and WTs under a large disturbance, it is found that IVS ability of an SM is generally stronger than a WTs due to its voltage source characteristic. Besides, the LVRT control-based var response of a WT is related to the TOV extent due to its control time delay. To meet the MRSCR and SVSCs in HWPPSs, a certain number of SMs should be reserved in the system to provide SCC and IVS. By solving the proposed UC model for multistage reactive power coordination, violations of the MRSCR and SVSCs can be eliminated and trip faults of WTs can be effectively avoided. The proposed solution method based on GBD can efficiently solve the MINLP model within an acceptable time and meet the requirements of day-ahead scheduling. With the application of new control strategies such as grid-forming control, the short-term voltage security issues and optimal operation in 100% IBG-penetrated power systems can be further studied.